\documentclass[preprint,amsmath,amssymb,aps,showkeys,showpacs]{revtex4}
\usepackage{graphicx}
\usepackage{graphics}
\usepackage{amsmath}
\usepackage{dcolumn}
\usepackage{amssymb}
\usepackage{bm}

\begin{document}
\title{Analysis of the interaction of an electron with radial electric fields in the presence of a disclination }
\author{K. Bakke}
\email{kbakke@fisica.ufpb.br}
\affiliation{Departamento de F\'isica, Universidade Federal da Para\'iba, Caixa Postal 5008, 58051-900, Jo\~ao Pessoa, PB, Brazil.}

\author{C. Furtado}
\email{furtado@fisica.ufpb.br}
\affiliation{Departamento de F\'isica, Universidade Federal da Para\'iba, Caixa Postal 5008, 58051-900, Jo\~ao Pessoa, PB, Brazil.}

\begin{abstract}

We consider an elastic medium with a disclination and investigate the topological effects on the interaction of a spinless electron with radial electric fields through the WKB (Wentzel, Kramers, Brillouin) approximation. We show how the centrifugal term of the radial equation must be modified due to the influence of the topological defect in order that the WKB approximation can be valid. Then, we search for bound states solutions from the interaction of a spinless electron with the electric field produced by this linear distribution of electric charges. In addition, we search for  bound states solutions from the interaction of a spinless electron with radial electric field produced by uniform electric charge distribution inside a long non-conductor cylinder.

\end{abstract}


\maketitle

\section{Introduction}

At present days, it is well-known that an elastic medium with a disclination can be described by the differential geometry \cite{kleinert,kat,val}. This way of describing the deformation of the elastic medium through a geometrical approach was proposed by Katanaev and Volovich \cite{kat}. With the Katanaev-Volovich proposal, the Volterra process \cite{kleinert}, which consists of the process of ``cut'' and ``glue'' of the elastic medium, was connected with the geometrical description of the a topological defect in the context of general relativity \cite{put}. Since topological defects can modify the electronic properties of the medium, then, studies of semiconductors and quantum dots in the presence of topological defects have attracted attention in the literature \cite{semi,semi3,semi4,au,furt,fur5}. Other works have dealt with topological defects in quantum rings \cite{fur6,fur3,dantas1}, electrons subject to the deformed Kratzer potential \cite{1}, electron gas in a cylindrical shell \cite{shell}, quantum holonomies \cite{bf2} and with an electron in a uniform magnetic field \cite{fur,fur2,fur7,fil}.   In recent years the influence of disclinations has been widely studied in several areas of physics\cite{furt}.  The presence of a disclination in medium modifies the physical properties of particles moving in it. Disclination in graphene layer , for example,  is formed from the removal or addition of an angular sector of the two-dimensional hexagonal lattice. In a recent work, Lammert and Crespi \cite{Lammert,Lammert2} used a  low-energy description of graphene layer  and started to study the electronic behavior of graphene monolayer with presence of disclination. Meanwhile,   Osipov, Kochetov and Pudlak \cite{osi1,osi2,osi3}  have investigated  the electronic structure disclination in graphene  using  analytical methods in self-consistent field-theory models .They have obtained the local and total density of states near the defects for these and other disclinated systems. Thus, it is very important to study the influence of disclination on particle dynamics in the presence of external electric and / or magnetic fields\cite{furt,furt2}

Since the beginning of the 20th century, the relationship between classical mechanics and quantum mechanics is the central theme of research in physics. In 1927, Sch\"odinger \cite{schro} found the wave equation that describes non-relativistic quantum systems. Shortly after, Wentzel \cite{wen}, Kramers \cite{kra} and Brillouin\cite{brill} developed a semi-classical approximation that has became popularly known in quantum mechanics as the WKB (Wentzel, Kramers, Brillouin) approximation \cite{griff,landau}. The WKB approximation gives an excellent description of conservative systems that have one degree of freedom and can be used to describe systems of varying degrees of freedom \cite{berryd}. In these systems, it can be demonstrated that its classic multidimensional dynamics can be decomposed in the dynamics of the corresponding number of one-dimensional conservative systems. Several authors have investigated alternative methods to improve the WKB approximation. An example is the modified WKB method proposed by Brozan \cite{brozan} that describes with good precision the ground state of nodeless systems with one more degrees of freedom. Another example is the proposal of Gomes {\it at al } \cite{gomes,gomes1,gomes2}, where an algebraic reformulation of the WKB method describes very well strongly confining potential. Recently, Friedrich and Trost \cite{frie} showed that the WKB wave functions can be used to obtain the precise or highly accurate results of quantum mechanics even in semi-classical limit. They have presented several applications which include the derivation of properties of bound and continuous states near the threshold of a potential. These properties are important for the understanding of many results observed in experiments with cold atoms.

In this work, we investigate the topological effects associated with a presence of a disclination in an elastic medium on the interaction of a spinless electron with radial electric fields. We search for bound states solutions to the Schr\"odinger equation via the semiclassical point of view of WKB (Wentzel, Kramers, Brillouin) approximation \cite{griff,landau,wkb4}. In particular, the geometrical approach of the disclination \cite{furt} shows that we need to work with the cylindrical symmetry. Therefore, we have to deal with WKB approximation by following the points raised by Langer \cite{wkb} and Berry and co-workers \cite{wkb2,wkb3}. Langer \cite{wkb} showed how to deal with the centrifugal term in the radial equation and the singularity at the origin in the spherical symmetry. This way of dealing with the centrifugal term and the singularity at the origin becomes known in the literature as the Langer modification or Langer transformation \cite{wkb2,wkb3,wkb4}. Since then, several works have been investigated with the WKB approximation in systems with spherical symmetry based on the Langer transformation \cite{wkb,wkb4,wkb10,wkb12,wkb13}. Later, Berry and co-workers \cite{wkb2,wkb3} showed the treatment of the WKB approximation with the cylindrical symmetry. It is worth pointing out that the WKB approximation has been used to analyse several systems, such as, anharmonic oscillators \cite{griff,wkb5,wkb8,wkb21}, inverse power-law potentials \cite{wkb6,wkb9,wkb15}, spin-orbit coupling \cite{wkb7}, polynomial potentials \cite{wkb11}, electron in a uniform magnetic field \cite{wkb14,wkb22}, $\mathcal{PT}$-symmetric quantum mechanics \cite{wkb16,wkb17,wkb18}, neutral particle with electric quadrupole moment \cite{b} and graphene \cite{wkb19,wkb20}.

The structure of this paper is: in section II, we introduce the line element that describes a disclination. Then, we analyse the interaction between a spinless electron with a radial electric field produced by a linear distribution of electric charges through WKB approximation by searching for bound state solutions; in section III, we analyse the interaction of a spinless electron with a radial electric field produced by a uniform volume distribution of electric charges inside a non-conductor cylinder; in section IV, we present our conclusions.

\section{electric field of a linear distribution of electric charges case}

Let us consider a long, thin, non-conductor cylinder with an inner radius $r_{0}$ that possesses a uniform distribution of electric charges on its symmetry axis. Let us assume that there exists a disclination inside the cylinder. The disclination is described by the line element \cite{kleinert,kat,furt}:
\begin{eqnarray}
ds^{2}=dr^{2}+\alpha^{2}\,r^{2}\,d\varphi^{2}+dz^{2},
\label{1.1}
\end{eqnarray}
where $0<\alpha<1$ is the parameter associated with the deficit of angle. Note that the azimuthal angle is defined in the range $0\leq\varphi\,\leq2\pi$.

Since the cylinder is long and thin, therefore, we can consider the uniform distribution of electric charges to be a linear distribution in the $z$-direction for $r>r_{0}$. Then, in the presence of the disclination, the electric field produced by this linear distribution of electric charges at $r\,>\,r_{0}$ is \cite{bf1}
\begin{eqnarray}
\vec{E}=\frac{\lambda}{\alpha\,r}\,\hat{r},
\label{1.2}
\end{eqnarray} 
where $\lambda>0$ is a constant associated with the linear distribution of electric charges. When the electron is placed in the region $r>r_{0}$, the potential energy that arises from the interaction of the electron with the radial electric field is 
\begin{eqnarray}
V\left(r\right)=\frac{\left|q\right|\,\lambda}{\alpha}\,\ln\left(r/r_{0}\right).
\label{1.4}
\end{eqnarray}
where we have taken $q=-\left|q\right|$ for the charge of the electron. Observe in Eqs. (\ref{1.2}) and (\ref{1.4}) the influence of the topological defect on the electric field and, as a consequence, on the potential energy,  since they depend on the parameter $\alpha$  that represents the influence of the topological defect in the medium. Hence, the time-independent Schr\"odinger equation for an electron that interacts with the radial electric field (\ref{1.2}), in the presence of the topological defect (\ref{1.1}), is written in the form:
\begin{eqnarray}
\mathcal{E}\psi=-\frac{\hbar^{2}}{2m}\,\left[\frac{\partial^{2}}{\partial r^{2}}+\frac{1}{r}\frac{\partial}{\partial r}+\frac{1}{\alpha^{2} r^{2}}\frac{\partial^{2}}{\partial\varphi^{2}}+\frac{\partial^{2}}{\partial z^{2}}\right]\psi+\frac{\left|q\right|\,\lambda}{\alpha}\,\ln\left(r/r_{0}\right)\,\psi.
\label{1.3}
\end{eqnarray}
Note that the operators $\hat{L}_{z}$ and $\hat{p}_{z}$ commute with the Hamiltonian operator given in the right-hand side of Eq. (\ref{1.3}), therefore, we can write the solution of the Schr\"odinger equation  (\ref{1.3}) in terms of the eigenvalues of these operators as follows:
\begin{eqnarray}
\psi\left(r,\,\varphi,\,z\right)=e^{il\varphi+ikz}\,R\left(r\right),
\label{1.5}
\end{eqnarray}
where $l=0,1,2,\ldots$ is the quantum number associated with the $z$-component of the angular momentum and $-\infty<\,k\,<\infty$ is the eigenvalue of the $z$-component of the linear momentum. Hence, by substituting (\ref{1.5}) into Eq. (\ref{1.4}), we obtain the following radial equation:
\begin{eqnarray}
R''+\frac{1}{r}R'-\frac{l^{2}}{\alpha^{2}r^{2}}R-\frac{2m\left|q\right|\,\lambda}{\hbar^{2}\alpha}\,\ln\left(r/r_{0}\right)R+\left[\frac{2m\mathcal{E}}{\hbar^{2}}\,-k^{2}\right]R=0.
\label{1.6a}
\end{eqnarray}
Next, let us take 
\begin{eqnarray}
R\left(r\right)=\frac{1}{\sqrt{r}}\,u\left(r\right),
\label{1.6}
\end{eqnarray}
and then, the radial equation becomes
\begin{eqnarray}
u''-\frac{\left(\frac{l^{2}}{\alpha^{2}}-1/4\right)}{r^{2}}u-\frac{2m\left|q\right|\,\lambda}{\hbar^{2}\alpha}\,\ln\left(r/r_{0}\right)u+\left[\frac{2m\mathcal{E}}{\hbar^{2}}-k^{2}\right]u=0.
\label{1.7}
\end{eqnarray}

It has been shown in Refs. \cite{wkb3,wkb4} that the WKB approximation is valid when the cylindrical symmetry is present if we replace $\left(l^{2}-1/4\right)$ with $l^{2}$ in the centrifugal term of the radial equation. In the present case, due to the topological effects of the disclination, we have to make the replacement:
\begin{eqnarray}
\left(\frac{l^{2}}{\alpha^{2}}-\frac{1}{4}\right)\rightarrow\frac{l^{2}}{\alpha^{2}}.
\label{eq:}
\end{eqnarray}
Then, we have in Eq. (\ref{1.7}):
\begin{eqnarray}
u''-\frac{l^{2}}{\alpha^{2}r^{2}}u-\frac{2m\left|q\right|\,\lambda}{\hbar^{2}\alpha}\,\ln\left(r/r_{0}\right)u+\left[\frac{2m\mathcal{E}}{\hbar^{2}}-k^{2}\right]u=0.
\label{1.7a}
\end{eqnarray}

Observe that we can reduce the system to the plane $z=0$ by taking $k=0$. In this way, we define
\begin{eqnarray}
Q\left(r\right)=\sqrt{2m\left[\mathcal{E}-\frac{\left|q\right|\,\lambda}{\alpha}\,\ln\left(r/r_{0}\right)\right]-\frac{l^{2}\hbar^{2}}{r^{2}}},
\label{1.8}
\end{eqnarray}
and thus, the radial equation (\ref{1.7a}) takes the form:
\begin{eqnarray}
u''+\frac{Q^{2}\left(r\right)}{\hbar^{2}}\,u=0.
\label{1.8}
\end{eqnarray}

With the WKB approach, the radial wave function is written in the form \cite{wkb3,wkb4}:
\begin{eqnarray}
u\left(r\right)\cong\frac{2}{\sqrt{Q\left(r\right)}}\,\cos\left(\frac{1}{\hbar}\,\int_{r_{1}}^{r}Q\left(r'\right)\,dr'-\frac{\pi}{4}\right).
\label{1.9}
\end{eqnarray}

Then, with the cylindrical symmetry, it has been shown in Refs. \cite{wkb3,wkb4} that the Bohr-Sommerfed quantization is given by 
\begin{eqnarray}
\frac{1}{\hbar}\,\int_{r_{1}}^{r_{2}}Q\left(r\right)\,dr=\left(n-\frac{1}{2}\right)\pi,
\label{1.10}
\end{eqnarray}
where $n=1,2,3,\ldots$ is the quantum number associated with the radial modes, and $r_{1}$ and $r_{2}$ are the turning points. It has been pointed out in Refs. \cite{wkb,wkb2,wkb3,wkb4} that the wave function (\ref{1.12}) is well-behaved at $r_{1}=0$ when we work with $s$ waves. Thereby, we have only one turning point determined when $\mathcal{E}=V\left(r_{2}\right)$. Note that the $s$ waves are defined by taking $l=0$. Therefore, for  the present case, we obtain
\begin{eqnarray}
r_{2}=r_{0}\,\exp\left(\frac{\alpha\,\mathcal{E}}{\left|q\right|\lambda}\right).
\label{1.11}
\end{eqnarray}

Thereby, by working with $s$ waves and by performing a change of variables $r=r_{2}\,e^{-x}$ (which is known as a Langer-type transformation \cite{wkb3,wkb4}), then, the left-hand side of Eq. (\ref{1.10}) becomes
\begin{eqnarray}
\frac{1}{\hbar}\,\int_{0}^{r_{2}}Q\left(r\right)\,dr&=&\frac{r_{2}}{\hbar}\,\sqrt{\frac{2m\left|q\right|\lambda}{\alpha}}\int_{0}^{\infty}\sqrt{x}\,\,e^{-x}\,dx\nonumber\\
[-2mm]\label{1.12}\\[-2mm]
&=&\frac{r_{2}}{\hbar}\,\sqrt{\frac{2m\left|q\right|\lambda}{\alpha}}\,\frac{\sqrt{\pi}}{2}.\nonumber
\end{eqnarray}

By substituting Eq. (\ref{1.12}) into Eq. (\ref{1.10}), we obtain the energy levels $\mathcal{E}_{n\,l,\,k}$:
\begin{eqnarray}
\mathcal{E}_{n,0,0}=\frac{\left|q\right|\lambda}{\alpha}\,\ln\left(\frac{\hbar}{r_{0}}\sqrt{\frac{2\pi\alpha}{m\left|q\right|\lambda}}\left[n-\frac{1}{2}\right],\right).
\label{1.13}
\end{eqnarray}
where $n=1,2,3,\ldots$ is the quantum number associated with the radial modes. Hence, through WKB approximation, we obtain a discrete spectrum of energy from the interaction between an electron and a radial electric field produced by a linear distribution of electric charges in the presence of a disclination. We can observe that the allowed energies (\ref{1.13}) depends on the parameter $\alpha$ that characterizes the linear topological defect. Since there is no interaction between the electron and the topological defect, therefore, the presence of the parameter $\alpha$ in the allowed energies yields an analogue of the Aharonov-Bohm effect for bound states \cite{pesk,ab}. By taking $\alpha\rightarrow1$, we obtain the energy levels in the absence of topological defect, i.e.,
\begin{eqnarray}
\mathcal{E}'_{n,0,0}=\left|q\right|\lambda\,\ln\left(\frac{\hbar}{r_{0}}\sqrt{\frac{2\pi}{m\left|q\right|\lambda}}\left[n-\frac{1}{2}\right]\right),
\label{1.14}
\end{eqnarray}

Finally, let us return to the energy levels (\ref{1.13}) and analyse the spacing between the energy levels. Then, we have that 
\begin{eqnarray}
\mathcal{E}_{n+1,0,0}-\mathcal{E}_{n,0,0}=\frac{\left|q\right|\lambda}{\alpha}\,\ln\left(\frac{n+1/2}{n-1/2}\right).
\label{1.15}
\end{eqnarray}

Hence, the spacing between the energy levels does not depend on the mass of the electron. However, there is the influence of the topological defect on it due to the presence of the parameter $\alpha$ that characterizes the disclination. This means that the topology of the defect determines the spacing between the energy levels. Note that by taking $\alpha\rightarrow1$ in Eq. (\ref{1.15}), we obtain the spacing between the energy levels in the absence of the topological defect.

\section{electric field of a uniform volume distribution of electric charges case}

In this section, we analyse the interaction of a spinless electron with a radial electric field produced by uniform electric charge distribution inside a long non-conductor cylinder in the presence of the disclination (\ref{1.1}). Hence, the electric field inside this long non-conductor cylinder in the presence of the disclination is given by \cite{bf3}
\begin{eqnarray}
\vec{E}=\frac{\rho\,r}{2}\,\hat{r},
\label{5.1}
\end{eqnarray}
where $\rho>0$ is a constant related to the uniform (positive) electric charge distribution. Since $q=-\left|q\right|$ for the electron, thus, the time-independent Schr\"odinger equation becomes
\begin{eqnarray}
\mathcal{E}\psi=-\frac{\hbar^{2}}{2m}\,\left[\frac{\partial^{2}}{\partial r^{2}}+\frac{1}{r}\frac{\partial}{\partial r}+\frac{1}{\alpha^{2}r^{2}}\frac{\partial^{2}}{\partial\varphi^{2}}+\frac{\partial^{2}}{\partial z^{2}}\right]\psi+\frac{\left|q\right|\rho}{4}\,r^{2}\,\psi
\label{5.2}
\end{eqnarray}

By following the steps from Eq. (\ref{1.5}) to Eq. (\ref{1.6}), we find the following radial equation:
\begin{eqnarray}
u''-\frac{\left(\frac{l^{2}}{\alpha^{2}}-1/4\right)}{r^{2}}\,u-\frac{\omega^{2}}{\hbar^{2}}\,r^{2}\,u+\frac{2m\mathcal{E}}{\hbar^{2}}\,u=0,
\label{5.3}
\end{eqnarray}
where we have defined the parameters 
\begin{eqnarray}
\omega^{2}=\frac{m\,\rho\,\left|q\right|}{2}.
\label{5.4}
\end{eqnarray}

Next, let us also reduce the system to the plane $z=0$ by taking $k=0$ and replace the term $\left(\frac{l^{2}}{\alpha^{2}}-1/4\right)$ with $l^{2}/\alpha^{2}$ as in the previous section. In this way, the radial equation (\ref{5.3}) becomes
\begin{eqnarray}
u''-\frac{l^{2}}{\alpha^{2}r^{2}}\,u-\frac{\omega^{2}}{\hbar^{2}}\,r^{2}\,u+\frac{2m\mathcal{E}}{\hbar^{2}}\,u=0.
\label{5.3a}
\end{eqnarray}

In contrast to the previous section, the potential energy does not depend on the parameter associated with the topological defect. On the other hand, we are able to analyse a case where bound states solutions can be achieved via WKB approximation for all values of the angular momentum quantum number $l$. The effects of the topology of the defects gives rise to an effective angular momentum quantum number $l_{\mathrm{eff}}=l/\alpha$ in the centrifugal term of the radial equation (\ref{5.3a}). This shift in the angular momentum quantum number corresponds to an analogue of the Aharonov-Bohm effect \cite{pesk,ab}, since there is no interaction between the quantum particle and the topological defect \cite{fur6,fur2}. Therefore, with the purpose of working with WKB approximation, let us follow the steps from Eq. (\ref{1.7a}) to Eq. (\ref{1.8}) and define:
\begin{eqnarray}
Q\left(r\right)=\sqrt{2m\mathcal{E}-\omega^{2}\,r^{2}-\frac{l^{2}\hbar^{2}}{\alpha^{2}r^{2}}}.
\label{5.5}
\end{eqnarray}

Then, by performing a change of variables $x=r^{2}$, the left-hand side of Eq. (\ref{1.10}) becomes 
\begin{eqnarray}
\frac{1}{\hbar}\,\int_{r_{1}}^{r_{2}}Q\left(r\right)\,dr&=&\frac{\omega}{\hbar}\int_{r_{1}}^{r_{2}}\frac{1}{r}\sqrt{-r^{4}+\frac{2m\mathcal{E}}{\omega^{2}}\,r^{2}-\frac{l^{2}\,\hbar^{2}}{\alpha^{2}\omega^{2}}}\,\,\,dr\nonumber\\
[-2mm]\label{5.6}\\[-2mm]
&=&\frac{\omega}{\hbar}\int_{x_{1}}^{x_{2}}\frac{1}{x}\sqrt{-x^{2}+\frac{2m\mathcal{E}}{\omega^{2}}\,x-\frac{l^{2}\,\hbar^{2}}{\alpha^{2}\omega^{2}}}\,\,dx,\nonumber
\end{eqnarray}
where the turning points are
\begin{eqnarray}
x_{1}&=&\frac{m\,\mathcal{E}}{\omega^{2}}-\frac{1}{2}\sqrt{\left(\frac{2m\mathcal{E}}{\omega^{2}}\right)^{2}-\frac{4\,l^{2}\,\hbar^{2}}{\alpha^{2}\omega^{2}}};\nonumber\\
[-2mm]\label{5.7}\\[-2mm]
x_{2}&=&\frac{m\mathcal{E}}{\omega^{2}}+\frac{1}{2}\sqrt{\left(\frac{2m\mathcal{E}}{\omega^{2}}\right)^{2}-\frac{4\,l^{2}\,\hbar^{2}}{\alpha^{2}\omega^{2}}}.\nonumber
\end{eqnarray}
In this way, we obtain in Eq. (\ref{5.6}):
\begin{eqnarray}
\frac{1}{\hbar}\,\int_{r_{1}}^{r_{2}}Q\left(r\right)\,dr=\frac{\omega\pi}{4\hbar}\left[\frac{2m\mathcal{E}}{\omega^{2}}-\frac{\hbar\left|l\right|}{\alpha\omega}\right].
\label{5.8}
\end{eqnarray}

Hence, by substituting Eq. (\ref{5.8}) into Eq. (\ref{1.10}), we can also obtain the following spectrum of energy:
\begin{eqnarray}
\mathcal{E}_{n,\,l}=\hbar\sqrt{\frac{2\left|q\right|\rho}{m}}\left[n+\frac{\left|l\right|}{2\alpha}-\frac{1}{2}\right],
\label{5.9}
\end{eqnarray}
where $n=1,2,3,\ldots$ is the quantum number associated with the radial modes. Thereby, by using WKB approximation, we have also obtained in discrete spectrum of energy from the interaction between an electron and a radial electric field. In this case, the electric field is produced by a uniform electric charge distribution inside a long non-conductor cylinder that possessed a disclination. Observe the dependence of the energy levels (\ref{5.9}) on the parameter $\alpha$ that characterizes the disclination (\ref{1.1}). Therefore, we also have an analogue of the Aharonov-Bohm effect for bound states \cite{pesk,ab}. Note that the energy levels (\ref{5.9}) are analogous to spectrum of energy of the harmonic oscillator. The ground state energy is given by $\mathcal{E}_{1,\,0}=\hbar\sqrt{\frac{\left|q\right|\rho}{2m}}$. In addition, the energy levels are non-degenerate and equally spaced, i.e.,
\begin{eqnarray}
\mathcal{E}_{n+1,\,l}-\mathcal{E}_{n,\,l}=\hbar\sqrt{\frac{2\left|q\right|\rho}{m}}.
\label{5.9b}
\end{eqnarray}

Besides, in the limit $\alpha\rightarrow1$, we obtain the energy levels in the absence of defect, i.e.,
\begin{eqnarray}
\mathcal{E}'_{n,\,l}=\hbar\sqrt{\frac{2\left|q\right|\rho}{m}}\left[n+\frac{\left|l\right|}{2}-\frac{1}{2}\right].
\label{5.10}
\end{eqnarray}

\section{conclusions}

We have analysed two particular cases of the interaction of a spinless electron with radial electric fields in an elastic medium with a disclination. This analysis has been made based on WKB approximation with the purpose of finding bound state solutions. By focusing on the radial equation, we have observed that the effects of the topology of the disclination give rise to an effective angular momentum quantum number $l_{\mathrm{eff}}=l/\alpha$. For this reason, the WKB approximation has became valid by replacing $\left(\frac{l^{2}}{\alpha^{2}}-\frac{1}{4}\right)$ in the centrifugal term of the radial equation with $l^{2}/\alpha^{2}$.

In the first case analysed, we have seen that the topology of the disclination modifies the electric field produced by a linear distribution of electric charges and, in turn, it also influences the potential energy of the system. Then, by applying WKB approximation, we have seen that bound states can be obtained for $s$ waves, where there is the influence of the topology of the disclination on the discrete spectrum of energy of the system. Due to the fact that there is no interaction between the electron and the topological defect, hence, this corresponds to an analogue of the Aharonov-Bohm effect for bound states \cite{pesk,ab}. Besides, we have observed that the spacing between the energy levels are determined by the topology of the defect.

In the second case analysed, we have seen no effect of the topology of the disclination on both electric field and potential energy of the system. On the other hand, the effects of the topology of the disclination has given rise to an effective angular momentum quantum number $l_{\mathrm{eff}}=l/\alpha$. By applying WKB approximation, we have seen that bound states can be obtained for all values of the angular momentum quantum number $l$. The spectrum of energy is discrete and depends on the effective angular momentum quantum number $l_{\mathrm{eff}}=l/\alpha$. Since no interaction between the electron and the topological defect exists, the presence of $l_{\mathrm{eff}}=l/\alpha$ in the energy levels gives rise to an analogue of the Aharonov-Bohm effect for bound states \cite{pesk,ab}.

In view of searching for Aharonov-Bohm-type effects through WKB approximation, this work draws attention to the possibility of analysing the influence of linear topological defects on the interaction of neutral particles with non-uniform magnetic fields \cite{lin,bs} and non-uniform electric fields \cite{er,b,lin2}.

\acknowledgments{The authors would like to thank the Brazilian agency CNPq for financial support.}

\end{document}